\documentclass[aps,twocolumn,prl,showpacs]{revtex4}
\usepackage{graphicx}
\usepackage{amssymb}
\begin{document}

\title{Quantum interference spectroscopy of RbHe exciplexes formed on helium nanodroplets}

\author{M. Mudrich}
\email{mudrich@physik.uni-freiburg.de} \affiliation{Physikalisches
Institut, Universit\"at Freiburg, 79104 Freiburg, Germany}
\author{G. Droppelmann}
\author{P. Claas}
\affiliation{Fakult\"at f\"ur Physik, Universit\"at Bielefeld, 33615
Bielefeld, Germany}
\author{C.P. Schulz}
\affiliation{Max-Born-Institut, Max-Born-Stra\ss e 2a, 12489 Berlin,
Germany}
\author{F. Stienkemeier}
\affiliation{Physikalisches Institut, Universit\"at Freiburg, 79104
Freiburg, Germany}

\date{\today}
\begin{abstract}
Femtosecond multiphoton pump-probe photoionization is applied to
helium nanodroplets doped with rubidium (Rb). The yield of Rb$^+$ ions features
pronounced quantum interference (QI) fringes demonstrating the coherence
of a superposition of electronic states on a time scale of tens of
picoseconds. Furthermore, we observe QI in
the yield of formed RbHe exciplex molecules. The quantum
interferogram allows to determine the vibrational structure of these
unstable molecules. From a sliced Fourier analysis one can
not only extract the population dynamics of vibrational states but
also follow their energetic evolution during the RbHe formation.
\end{abstract}


\maketitle

One of the great achievements of femtosecond (fs) lasers has been the observation in real
time of the formation and breaking of chemical bonds between two
atoms~\cite{Zewail:1994}. The molecular dynamics is visualized when initializing a
non-stationary multi-state superposition (wave packet, WP) by a fs pump pulse, by letting
the WP evolve freely in time, and by projecting it onto a final state by the second probe
pulse (pump-probe (PP) technique). This final state is subsequently detected with
time-independent methods, e.g. measuring the spontaneous fluorescence or the yield of
photo ions.

A different approach to molecular WP dynamics is WP interferometry
based on excitation by two identical fs pulses with a well-defined
relative phase in an interferometric setup. This approach relies on
the interference of WP amplitudes excited according to two
temporally distinct quantum paths leading to the same final state
\cite{Brumer:2003}. In the limit of weak fields this approach is
equivalent to quantum beats, Ramsey fringes in the time domain
\cite{Noordam:1992,Christian:1993}, Fourier spectroscopy using fs
pulses \cite{Bellini:1997}, or temporal coherent control
\cite{Blanchet:1997}. The measured quantum interferograms carry the
high-frequency oscillation of the electronic energy modulated by the
low-frequency beatings of WP motion.

Direct measurement of QI oscillations has allowed to study the dynamics
of atomic Rydberg states, electronic spin and nuclear spin WPs of free atoms in the gas
phase
\cite{Noordam:1992,Christian:1993,Jones:1995,Bellini:1997,Blanchet:1997,Praekelt:2004},
of atoms on surfaces \cite{Ogawa:1999}, and of atoms attached to helium nanodroplets
\cite{Stienkemeier2:1999}. Different variants of WP interferometry have also been applied
to simple molecules
\cite{Scherer:1991,Blanchet:1998,Warmuth:2000,Ohmori:2003,Fushitani:2005} and even to
molecular crystals \cite{Tortschanoff:1999}. Using pump-probe spectroscopy with
phase-locked pulses the dynamics of electronic as well as vibrational coherence of Cl$_2$
molecules embedded in solid Ar has been investigated \cite{Fushitani:2005}. As a recent
highlight, high-precision WP interferometry with HgAr and I$_2$ dimers has been
demonstrated allowing to prepare arbitrary relative populations in different vibrational
states \cite{Ohmori:2003}.

Helium nanodroplets are widely applied as a nearly ideal cryogenic matrix for
spectroscopy of embedded molecules and as nanoscopic reactors for studying chemical
reactions at extremely low temperatures
\cite{Stienkemeier:2001,Toennies:2004,Lugovoj:2000}. Only recently, the real-time
dynamics of doped helium nanodroplets have been studied in pump-probe experiments
\cite{Stienkemeier:2006}. Alkali atoms and molecules represent a peculiar class of dopant
particles due to their extremely weak binding to the surface of helium nanodroplets. Thus
they can be viewed as intermediate systems between the gas phase and conventional
cryogenic matrices.

In this Letter, we report on QI spectroscopy upon
excitation of Rb atoms attached to helium nanodroplets. In
particular, we interpret interference structures in RbHe exciplex
molecules which are formed upon excitation of Rb into the $5p$
state, demonstrating that coherence even survives the formation of a
chemical bond. The quantum interferogram is analyzed to extract the
vibrational spectrum of the unstable RbHe molecule by Fourier
transformation.

In the experiment, a fs laser system is combined with a helium
nanodroplet molecular beam machine. The experimental setup is
identical to the one used in former experiments
\cite{Droppelmann:2004}. Superfluid helium nanodroplets are formed
in a supersonic expansion of helium gas at a high stagnation
pressure ($50\cdot10^5$\,Pa) from a cryogenic nozzle (T = 19\,K).
The generated droplets have a mean droplet size of 10\,nm and cool
by natural evaporation to a terminal temperature of 380\,mK which is
well below the transition temperature to superfluidity. The droplets are doped
with single alkali atoms which are picked up in a heated vapor cell
further downstream. The weak interaction of alkali atoms with helium
leads to so called ``bubbles'' where the solvation environment is
characterized by a diminished helium density. In helium droplets
alkalis therefore reside in surface states, i.e.~the atoms are
weakly bound on top of dimple-like textures
\cite{Stienkemeier:2004}. Because binding energies are only around
10\,K which is small compared to energies of laser-induced
processes, desorption from the droplets dominantly follows laser
excitations.

Pairs of fs laser pulses with $\approx 110$\,fs pulse duration and
equal intensity are generated by a Mach-Zehnder interferometer and
propagate collinearly with variable delay time up to 100\,ps with a
step increment down to 220 attoseconds. In the probe step Rb$^+$ and
RbHe$^+$ photoions are formed which are subsequently detected mass
selectively in a quadrupole mass spectrometer.

\begin{figure}
\resizebox{0.85\columnwidth}{!}{\includegraphics{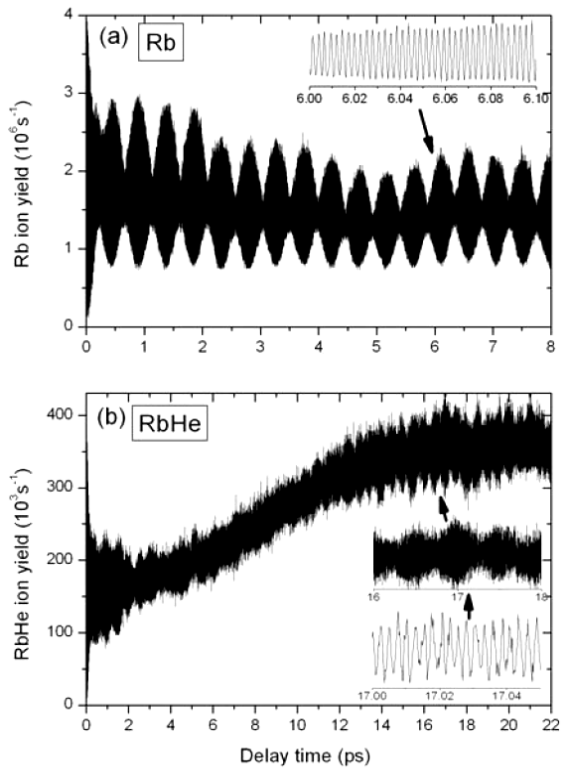}}
\caption{\label{fig:pptrace} (a) Yield of photo-ionized Rb atoms excited on the
surface of helium nanodroplets as a function of the delay between femtosecond pump and
probe pulses. The inset shows that by choosing a step increment of 220 attoseconds the
QI oscillation is fully resolved. (b) QI fringes of
ionized RbHe exciplexes detected on mass 89\,amu. The amplitude modulation comprehends
the vibrational structure of the RbHe molecule.}
\end{figure}

Note that Rb atoms are excited on the helium droplets but are detected
either as free atoms or as free RbHe exciplexes after
having left the droplet. The photo ionization signals detected on the mass of bare Rb$^+$
ions (85\,amu) and on the mass of RbHe$^+$ ions (89\,amu) as a function of delay time
between pump and probe pulse are depicted in Fig.~\ref{fig:pptrace} (a) and (b),
respectively. In (a), the laser wave length is tuned to $\lambda=775$\,nm which leads to
a $\Sigma$-type excitation with purely repulsive interaction between the excited
$p$-orbital of Rb and the He surface. In this case the Rb atoms rapidly desorb from the
He droplets. Consequently, the influence of the helium environment does not play a role
when probing free Rb atoms after long delay times.

\begin{figure}
\resizebox{0.8\columnwidth}{!}{\includegraphics{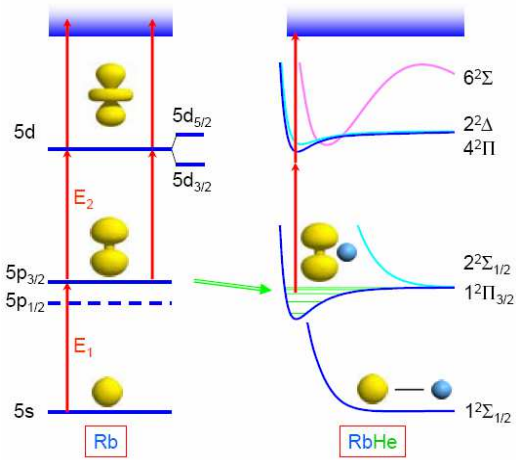}}
\caption{\label{fig:levels} Pump-probe excitation schemes showing the involved electronic
states of Rb atoms (left side) as well as the vibronic states after having formed a RbHe
exciplex~\cite{Pascale:1983}. The spatial distributions of the electron orbitals in corresponding states are
also sketched.}
\end{figure}

The ion yield is strongly modulated by QI oscillations. The fast
oscillation, corresponding to the excitation frequency with a period of about 2.6\,fs
(inset of Fig.~\ref{fig:pptrace} (a)), clearly displays a beat note with a period of
0.47\,ps. In addition, half a period of a 11\,ps-beat note is recognizable from the slow
drop of oscillation amplitude from $t=0$ until $t\approx 5$\,ps. These beat frequencies
are readily interpreted by energy differences between levels which are coherently excited
in the PP scheme, as illustrated in Fig.~\ref{fig:levels} (a). Coincidentally, the
5p$_{3/2}\leftarrow$ 5s ($E_1$ in Fig.~\ref{fig:levels}) and the 5d$_{3/2}$~(5d$_{5/2}$)
$\leftarrow$ 5p$_{3/2}$ ($E_2$ in Fig.~\ref{fig:levels}) transitions are energetically
different only by $\Delta E = E_2 - E_1 = 67.44$ $(70.40)$\,cm$^{-1}$, which is less than
the spectral width of the laser pulse (140\,cm$^{-1}$). Thus the two levels 5p and 5d can
be coherently excited by the pump pulse which causes the 0.47\,ps-beat pattern. The slow
11\,ps beat results from the 2.96\,cm$^{-1}$ fine structure splitting between 5d$_{3/2}$
and 5d$_{5/2}$ states. On the other hand, the $J=1/2$ fine structure component of the 5p
state is separated from the $J=3/2$ state by 237.6\,cm$^{-1}$ and does not contribute to
the spectra shown here.

At certain excitation wavelengths, where the alignment of the excited
$p$-orbital is parallel to the surface of the droplet, the
attractive interaction of this configuration leads to the formation
of RbHe exciplex molecules \cite{Droppelmann:2004}. The maximum
yield of RbHe exciplexes is obtained when tuning the fs laser to
$\lambda=780$\,nm. In previous PP experiments the formation time of
RbHe exciplexes has been measured to $\tau\approx 8$\,ps
\cite{Droppelmann:2004}. Accordingly, the RbHe$^+$ ion yield
depicted in Fig.~\ref{fig:pptrace} (b) increases continuously from
$t=0$ to $t=15$\,ps. The initial signal level of about 45\% of the
maximum signal is due to ionization of RbHe molecules which have been excited by previous laser pulse pairs
of the fs pulse train. Besides the exciplex formation dynamics, the PP transients recorded on the RbHe mass feature QI fringes as well~\cite{Droppelmann:2004}. Hence,
coherence of the electronic superposition state excited by the pump
pulse in Rb atoms appears to be conserved upon binding of a He atom
to the Rb. The amplitude of the RbHe QI also shows
a well defined modulation pattern. However, at first sight no
obvious beat frequencies can be identified, suggesting that a
variety of frequency components contribute to the beat signal.

\begin{figure}
\resizebox{1.05\columnwidth}{!}{\includegraphics{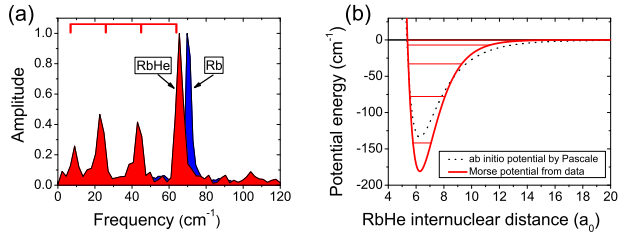}}
\caption{\label{fig:fft} a) Fourier transforms of the amplitude functions of the quantum
interferograms of the Rb and RbHe data shown in Fig.~\ref{fig:pptrace}. The spectrum of
Rb is dominated by the frequency component $E_2-E_1$,
%$E_{5p-5d}-E_{5s-5d}$
whereas the RbHe spectrum displays
a vibrational progression. b) Morse potential reconstructed from a fit of the vibrational
spectrum in comparison with an \textit{ab initio} potential \cite{Pascale:1983}.
}
\end{figure}

In this Letter, the interference fringes are quantitatively analyzed for extracting
information about the coherently excited energy levels. To this end,
the PP transients of Fig.~\ref{fig:pptrace} are Fourier transformed in the entire
delay time range. The Fourier spectra of the envelope function of the interference
oscillations are shown in Fig.~\ref{fig:fft} a). As expected, the Rb spectrum features
only one frequency component corresponding to the energy difference $E_2-E_1$. The
observed value $70.4(3)$\,cm$^{-1}$ matches the value for the 5d$_{5/2}$ $\leftarrow$
5p$_{3/2}$ transition, due to the larger transition moment to the 5d$_{5/2}$ state
compared with the 5d$_{3/2}$ state.

The RbHe spectrum contains 4 distinct maxima at frequencies $8.8(6)$\,cm$^{-1}$,
$23.2(7)$\,cm$^{-1}$, $43.7(2)$\,cm$^{-1}$, and at $65.8(3)$\,cm$^{-1}$. These
frequencies can be attributed to the energy differences between neighboring vibrational
levels of the RbHe molecule. Because of the unstable nature of the molecule, these have
not been determined experimentally before. Therefore only theoretical predictions are
available for comparison. We constructed a Morse potential from a linear fit of the
measured level spacings. The fit result is included in Fig.~\ref{fig:fft} as a comb
representation. The corresponding Morse potential is depicted by the solid line in
Fig.~\ref{fig:fft} b). For comparison, the \textit{ab initio} potential of Pascale
\cite{Pascale:1983} is plotted in the same diagram as a dashed line. The position of the
minimum of the Morse potential is set to coincide with the one of the \textit{ab initio}
potential.
%The depth of the well of the Morse potential (-180.9\,cm$^{-1}$), which is
%determined by the vibrational level spacings, agrees to better than 5\,\% with the one of
%the \textit{ab initio} potential.
The well depths of the Morse and the \textit{ab initio} potentials (-180.9\,cm$^{-1}$ and -133.9\,cm$^{-1}$, respectively)
deviate by 25\%, which is reasonable agreement in view of the limited accuracy of the
\textit{ab initio} calculations and the simple model potential used to fit the data.
%The shapes of the two potential curves are
%quite different due to the simple model potential used to fit the data and probably due
%to the impreciseness of the \textit{ab initio} potential.

The relative heights of
measured peaks indicate that all vibrational levels are populated with similar
probabilities. This goes perfectly along with observations of emission spectra in KHe
exciplexes for which the populations of vibrational states were quantified
\cite{Reho:2000}: Depending on the excitation laser frequency, slightly more pronounced
lower or higher vibrational states, respectively, have been observed in this very similar
system. In our experiment we can also tune the population of vibrational levels by
varying the femtosecond laser frequency or pulse characteristics.

\begin{figure}
\includegraphics[scale=1.0]{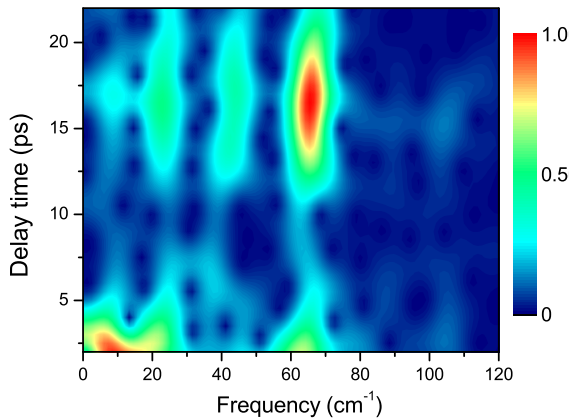}
\caption{\label{fig:spec} Spectrogram of the RbHe QI oscillation
showing the time evolution of the energy and the population of the coherently excited
states. }
\end{figure}

More importantly, it is shown that our real-time approach not only
unravels the vibrational energy structure of the unstable RbHe
molecule but that we can even follow the evolution of the
vibrational structure during the formation process of the molecule.
The time dependence of the frequency components can be visualized by
Fourier transforming the data inside a time window of fixed length
(2\,ps) which slides across the entire scan range. The resulting
spectrogram of RbHe is depicted in Fig.~\ref{fig:spec} as a contour
plot.
%The individual frequency components discussed
%before are identified in the spectrogram. In the upper half of the
%plot at delay times $t\geq 12\,$ps they appear as vertical lines,
%meaning that energies do not change with time.
The individual frequency components discussed
before appear as vertical lines in the upper half of the spectrogram at delay times $t\geq 12\,$ps,
meaning that energies do not change with time.
In contrast to that, at short delay times ($<10$\,ps) one can observe shifts. These are not very pronounced in the data set presented in this paper but are clearly present when using different excitation wavelengths (not shown in this paper). The shifts are of the order 10\,cm$^{-1}$, which nicely correlates
with the binding energy of RbHe to the droplet prior to desorption.

%In contrast to that in the lower half the maxima do shift. This effect is even more
%pronounced at different excitation wavelengths (not shown in this
%paper). The shifts are about 10\,cm$^{-1}$, which nicely correlates
%with the binding energy to the droplet.
In the formation process the molecules are formed near the dissociation limit
and then vibrationally cool through the interaction with the helium droplet before they desorb and
subsequently keep their population~\cite{Reho:2000}. This cooling mechanism can also be followed in the spectrogram:
At delay times shorter than 4\,ps there is much more intensity at lower frequencies which
represent energy differences between higher lying levels. At longer times, intensity distributes
almost equally over all the lines. By changing the excitation laser wave length by a few nm,
the relative amplitudes of the individual lines can be slightly changed.

Finally, the issue of decoherence has to be discussed. It is striking that during the
observed range of delay times (20\,ps) the amplitude of the interference fringes
does not decrease at all, although initially the phase information is put into the
excitation of the very weakly bound Rb atoms on the droplets and the oscillation is
observed in the yield of RbHe molecules which have formed and detached from the droplet.
%It is important to recall that tunneling into the diatomic conformation was proposed as
%mechanism to form the RbHe bond~\cite{Reho:2000}.
% Quantum mechanical tunneling is a
%coherent process and describes a coupled system in which the probability of finding an
%objects at behind the barrier rises in time. Hence tunneling in itself does not introduce
%any decoherence into the system. ]
As discussed by Shapiro~\cite{Shapiro:2006}, in dissociation reactions coherence
of the fragments is maintained in the case of broadband laser excitation. The same
arguments can be applied to association of RbHe molecules. However, cooling of
vibrational excitations of the RbHe molecule by dissipation of energy into the helium
droplet must be allowed for as a source of decoherence. Indeed, in the case of
KHe exciplex formation, a fast decay of the QI contrast within 1.5\,ps
is observed~\cite{Stienkemeier2:1999}. Thus, the dynamics of decoherence in these systems
that are weakly coupled to helium nanodroplets appears to sensitively depend on the
internal structure and on the coupling strength to the helium environment. This will be
studied more systematically in future experiments.
%[ However, decoherence delicately depends on the structure, the coupling
%and population of states and coherent control has been successfully applied despite
%collisional effects~\cite{Brumer:2003}. After desorption of the RbHe exciplex from the He
%droplet one expects the amplitude only to undergo dephasing and rephasing and not to
%decrease further. ]
The result of our experiments that electronic coherence
survives the formation and cooling process without severe suppression is a valuable piece
of information, demonstrating that coherent control techniques are not impeded by the
formation or breaking of bonds. In comparison with large, e.g.~biological molecular
complexes, our system is, of course, a clean and well defined process which relies on
such low temperatures.

In conclusion, we have demonstrated that QI structures can be used not
only to determine binding properties of small molecules but also to characterize the
reaction process in real time. Changes in the vibrational structure as well as changes in
their population can be directly monitored. More importantly, the bond formation process
does not destroy coherent properties showing that this kind of coherent control
experiments can in principle be applied to more complex systems.

Stimulating discussion with Moshe Shapiro and financial support by the Deutsche Forschungsgemeinschaft are gratefully acknowledged.

\end{document}